# Computing Pure Nash Equilibria via Markov Random Fields

Constantinos Daskalakis*

June 19, 2018

**Abstract**

In this paper we present a novel generic mapping between Graphical Games and Markov Random Fields so that pure Nash equilibria in the former can be found by statistical inference on the latter. Thus, the problem of deciding whether a graphical game has a pure Nash equilibrium, a well-known intractable problem, can be attacked by well-established algorithms such as Belief Propagation, Junction Trees, Markov Chain Monte Carlo and Simulated Annealing. Large classes of graphical games become thus tractable, including all classes already known, but also new classes such as the games with $O(\log n)$ treewidth.

## 1 Introduction

Games require for their description data in general exponential in the number of players. Indeed, a normal form game of $n$ players and $s$ strategies available to each player needs $ns^n$ numbers to be described. This is because it is assumed that every player directly depends on the strategy of every other player. This exponential dependency of the description size on the number of players is, obviously, forbidding for the study of games with a large number of players. Furthermore, such exponential complexity may not be necessary: for many game domains of interest, such as markets and the Internet, it can be argued that the welfare of a player depends directly on only a few other players. This observation allows for much more succinct representations of games which exploit the dependencies between players more explicitly than does the classical representation.

One important class of succinct games is that of *graphical games*, which was suggested by Kearns et al. [KLS01]. In a graphical game, we are given a graph with the players as nodes. It is postulated that a player's utility depends on the strategy chosen by the player *and by the player's neighbors in the graph*. Such games played on graphs of bounded degree can be represented by polynomially many (in $n$ and $s$) numbers. Graphical games are quite attractive as models of the interaction of agents across a large network or market. There has been a host of positive complexity results for this kind of games. It has been shown, for example, that *correlated equilibria* (a sophisticated equilibrium concept suggested by Aumann [Aum74]) can be computed in polynomial time for graphical games that are trees [KKLO03], which was later extended to all graphical games [Pap05]. Moreover, it has been shown that, in some cases, even mixed Nash equilibria can be computed efficiently [LKS01].

From the famous theorem by Nash [Nas51] it follows that both correlated and mixed Nash equilibria are guaranteed to exist in every game. The same does not hold for *pure Nash equilibria* — the deterministic counterparts of mixed Nash equilibria. Pure Nash equilibria are not guaranteed to exist and, in fact, deciding whether a graphical game has a *pure Nash equilibrium* is an NP-complete problem. *NP-completeness holds*

---

*UC Berkeley, Computer Science Division, Soda Hall, Berkeley, CA 94720. Email: costis@cs.berkeley.edu.

*even when restricted to the class of graphical games defined on bipartite graphs of degree bounded by 3 and 3 strategies available to each player* (see e.g. [GGS03]). Therefore, a reasonable question is whether there exist large classes of graphical games for which deciding the existence of pure Nash equilibria and computing one or all pure Nash equilibria can be done efficiently. Moreover, is it possible to design algorithms that perform well on general hard instances of the problem? These questions are the focus of the present paper.

Besides the economy of description, one of the motivations for the introduction of graphical games was the intuitive affinity between graphical games and graphical statistical models; indeed, several algorithms for graphical games (e.g., [LKS01, OK02]) do have the flavor of algorithms for Bayes nets. However, a direct connection between graphical games and graphical statistical models had not been made explicit — certainly not in the context of pure Nash equilibria.

In this paper we present a mapping from any graphical game to a Markov Random Field (MRF) with the following properties:

- Finding a Maximum-A-Posteriori configuration of the MRF answers the question of whether the graphical game has a pure Nash equilibrium.

- The marginal probability distributions of the cliques of the MRF constitute a succinct description of all pure Nash equilibria of the graphical game. Note that there might be exponentially many pure Nash equilibria so we might not be able to compute all of them explicitly in input polynomial time.

- Sampling the distribution of the MRF provides a randomized algorithm for testing whether a graphical game has a pure Nash equilibrium and for computing pure Nash equilibria.

As a consequence, any statistical inference algorithm (from Belief Propagation to Markov Chain Monte Carlo methods [MRR$^+$53] and Simulated Annealing [KGV87], see sections 4 and 5) can be used to compute pure Nash equilibria. Moreover, by combining this mapping with the Junction Tree Algorithm [LS90, JLO90] we show that for large classes of graphical games we can compute in polynomial time a succinct description of the set of pure Nash equilibria (including, among new results, all previously known efficient algorithms for pure Nash equilibria [GGS03]):

- Graphical Games of bounded treewidth

- Graphical Games of bounded hypertree width*

- Graphical Games of $O(\log n)$-treewidth, bounded neighborhood size and bounded cardinality strategy sets

We believe that the latter class of graphical games is of special interest as a plausible model of networked markets. In such games, bounded neighborhood size and bounded number of strategies is a realistic assumption (and, to some extent, essential for concise representation of the game), but the number of players can be very large, while the game graph has a rich cycle structure. In fact, our result for this class of graphical games is the first positive result that is not based on some assumption about the cycle structure of the graph. Moreover, given that NP-completeness of computing pure Nash equilibria holds even when restricted to the class of games defined on graphs of degree 3 and at most 3 strategies available to each player, our result is quite tight. For an alternative way to approach large games, exploiting the periodic structure of the graph, see [DP05].

The structure of the paper is as follows. In section 2 we give the basic definitions and in section 3 we describe our reduction from Graphical Games to Markov Random Fields. In section 4 we describe how the reduction can be combined with the Junction Tree algorithm to yield polynomial time pure Nash equilibria algorithms for large classes of graphical games. Finally, in section 5 we suggest a deterministic algorithm for the general case and we describe how randomized algorithms can be derived using Markov Chain Monte

---

*Associated with every graphical game is a hypergraph as described in section 2.1; hypertree width is a measure of the degree of cyclicity of hypergraphs [GLS02] and its formal definition is given in section A of the appendix.

Carlo methods and Simulated Annealing. We, also, suggest the use of Survey Propagation [BMZ02] for solving hard instances of the problem.

## 2 Preliminaries

### 2.1 Graphical Games

In a *game* we have *n players*, $1, \ldots, n$. Each player $p$, $1 \leq p \leq n$, has a finite set of *strategies* or *choices*, $S_p$, with $|S_p| \geq 2$, and a *payoff* function $u_p : \prod_{i=1}^n S_i \to \mathbb{N}$. The set $S = \prod_{i=1}^n S_i$ is called *set of strategy profiles* and we denote the set $\prod_{i \neq p} S_i$ by $S_{-p}$.

It's clear that, in order to specify a game with $n$ players and $s$ strategies each, we need $ns^n$ numbers, an amount of information exponential in the number of players. However, players often interact with a limited number of other players, and this allows for much more succinct representations:

**Definition 2.1** A *graphical game* $\mathcal{G} = \langle G, \{S_p\}, \{u_p\}\rangle$ is defined by:
- An undirected graph $G = (V, E)$, where $V = \{1, \ldots, n\}$ is the set of players.
- For every player $p \in V$:
    - A non-empty finite set of *strategies* $S_p$
    - A *payoff function* $u_p : \prod_{i \in \mathcal{N}(p)} S_i \to \mathbb{N}$ (where $\mathcal{N}(p) = \{p\} \cup \{v \in V | (p, v) \in E\}$)

We note that a different perspective from which we can see graphical games, which will be useful later, is through the *hypergraph* they induce. We can imagine that every game $\mathcal{G}$ defines a hypergraph having as nodes the players and, for every player $p$, one hyperedge containing $p$'s neighborhood, $\mathcal{N}(p)$; suppose that we remove duplicate hyperedges if two or more players have the same neighborhood. We denote this hypergraph by $\mathcal{H}(\mathcal{G})$ and we define as *primal graph of the game* the primal graph[†] of $\mathcal{H}(\mathcal{G})$.

### 2.2 Pure Nash Equilibrium - Best Response Function

Consider a game with $n$ players and strategy sets $S_1, \ldots, S_n$. For every strategy profile $s \in S$, we denote by $s_p$ the strategy of player $p$ in this strategy profile and by $s_{-p}$ the $(n-1)$-tuple of strategies of all players but $p$. For every $s'_p \in S_p$ and $s_{-p} \in S_{-p}$ we denote by $(s_{-p}; s'_p)$ the strategy profile in which player $p$ plays $s'_p$ and all the other players play according to $s_{-p}$.

**Definition 2.2 (Pure Nash Equilibrium)**
A strategy profile $s$ is a *pure Nash equilibrium* if for every player $p$ and strategy $t_p \in S_p$ we have $u_p(s) \geq u_p(s_{-p}; t_p)$.

Intuitively, a strategy profile $s$ is a pure Nash equilibrium if none of the players has a unilateral incentive to deviate: the player cannot increase his/her payoff by deviating to another strategy if the other players continue to play the strategies in $s_{-p}$.

**Definition 2.3 (Best Response Function)**
The *Best Response Function of player* $p$ is a function $\mathrm{BR}_{u_p} : S_{-p} \to 2^{S_p}$ defined by:

$$\mathrm{BR}_{u_p}(s_{-p}) \triangleq \{s_p | s_p \in S_p \text{ and } \forall s'_p \in S_p : u_p(s_{-p}; s_p) \geq u_p(s_{-p}; s'_p)\}$$

---

[†]The primal graph $G' = (V', E')$ of a hypergraph $\mathcal{H} = (\mathcal{V}, \mathcal{E})$ has $V' = \mathcal{V}$ and two nodes $v_1, v_2 \in V'$ are connected iff there is a hyperedge $h \in \mathcal{E}$ such that $v_1, v_2 \in h$.

Intuitively, $\text{BR}_{u_p}(s_{-p})$ is the set of strategies in $S_p$ that maximize $p$'s payoff if the other players play $s_{-p}$.

It's easy to see that we can define pure Nash equilibrium in terms of the best response functions of players. Indeed, a strategy profile $s$ is a pure Nash equilibrium if, for every player $p$, $s_p \in \text{BR}_{u_p}(s_{-p})$.

## 2.3 Markov Random Fields and Statistical Inference

*MRFs:* We describe informally the notion of an undirected graphical model and we refer the reader to [Lau96] for a more detailed description. An ***Undirected Graphical Model***, or ***Markov Random Field***, over an undirected graph $G = (V, E), |V| = n$, is a probability distribution that factorizes according to functions defined on a set $\mathcal{C}$ of cliques of $G$[‡]. More precisely, associated with every node $v \in V$ is a random variable $x_v$ taking values from a set $\mathcal{X}_v$ of values[§]. Also, associated with every clique $c \in \mathcal{C}$ is a potential function $\psi_c : \prod_{v \in c} \mathcal{X}_v \to \mathbb{R}_+$ that depends only on $x_c = \{x_v | v \in c\}$. Using this notation the probability distribution defined on $\mathbf{x} = \{x_v | v \in V\}$ is:

$$p(\mathbf{x}) = \frac{1}{Z} \prod_{c \in \mathcal{C}} \psi_c(x_c) \tag{1}$$

where $Z$ is a normalizing constant. We'll refer to $\mathcal{C}$ as the *set of significant cliques of the MRF*.

*The principal inference problems:* Some of the principal statistical inference problems defined on Markov Random Fields are the following and the literature that addresses them is very rich.

1. *Maximum-A-Posteriori (MAP) Estimation:* the problem of finding a configuration that is more likely under the distribution $p(\mathbf{x})$, or, more formally, of finding a configuration $\hat{x}_{MAP} \in \arg\max_{x \in \prod_v \mathcal{X}_v} p(\mathbf{x})$.

2. *Computing the marginal probability distribution* of a particular subset of the nodes or some subsets of the nodes simultaneously; usually the marginal probability distributions of the cliques of set $\mathcal{C}$.

3. *Sampling* the distribution $p(\mathbf{x})$.

*A crucial observation about the normalizing constant Z:* In order to compute $Z$ one has to sum $p(x)$ over all configurations $x \in \mathcal{X} = \prod_{v \in V} \mathcal{X}_v$, a computation that would require exponential time in the number of the nodes. However, this is not really needed for the above inference problems. Indeed, computing a MAP configuration does not change whether we include $Z$ in the computation or not. Also, sampling the distribution is usually based on the ratio of probabilities of configurations (e.g. in the Metropolis-Hastings sampling method) and so $Z$ is cancelled. Finally, computing marginal probability distributions involves summing $p(x)$; this can be done without including constant $Z$ and the resulting function can be normalized after the completion of the algorithm, since the marginal distribution is a probability distribution and thus must be normalized; now, since the marginal is usually computed on a subset of few nodes, the time needed to normalize the computed function is much less than that required to find $Z$. Thus, a very common practice in statistical inference is to assume **Z=1** for all computations. This is a key assumption as will become clear later (see discussion in section 3).

## 2.4 Clique Trees, Treewidth and the Junction Tree Algorithm

The Junction Tree Algorithm is one of the most celebrated algorithms for statistical inference and is used both for computing marginal distributions and for computing maximum-a-posteriori configurations. It is also in the core of numerous other algorithms for MRFs and Bayesian Networks. For a quick description of the algorithm we refer the reader to [WJ03]; a detailed description can be found in [JLO90, LS90]. Here we describe briefly some ingredients of the algorithm. Before doing so, let's define the notion of a clique tree of a graph, which is an equivalent way of defining a tree decomposition as implied by lemma 2.6.

---

[‡]Usually $\mathcal{C}$ is chosen to be the set of maximal cliques of graph $G$, but in some cases it is a different set. The choice depends on the underlying application.

[§]All Markov Random Fields that we consider in this paper will have finite sets $\mathcal{X}_v$. We will assume that this is the case in the rest of the paper.

**Definition 2.4** A *clique tree of a chordal graph* $G = (V, E)$ is a tree $T = (\mathcal{C}, \mathcal{E})$, where $\mathcal{C}$ is a subset of the set of all cliques of graph $G$, that has the following properties:
- $\forall$ clique $c$ of $G$, $\exists c' \in \mathcal{C}$ s.t. $c \subseteq c'$ (thus all maximal cliques of $G$ are nodes of $T$)
- $\forall c_1, c_2 \in \mathcal{C}, \forall c_3 \in \mathcal{C}$ in the unique path between $c_1$ and $c_2$, $c_1 \cap c_2 \subseteq c_3$ (clique intersection property)

The **width** of $T$ is $max_{c \in \mathcal{C}}\{|c|\}$.

**Definition 2.5** A *clique tree of a graph* $G = (V, E)$ is a clique tree of some triangulation of $G$.

**Lemma 2.6 (e.g. [Klo94])** *Every clique tree of a graph $G$ is a tree decomposition and vice versa. Thus, the treewidth of a graph $G$ is equal to the minimum width of all clique trees of $G$ minus 1.*

Briefly, the Junction Tree Algorithm does the following. It starts from the graph $G = (V, E)$ of the MRF and triangulates it to get a chordal graph $G' = (V, E')$. Then it builds a clique tree $T = (\mathcal{C}, \mathcal{E})$ of $G'$ and it loads to every node of $T$ a potential function as follows: it assigns the potential function of every significant clique of the MRF to exactly one of the nodes of $T$ that contain it and then for every node of $T$ it takes the product of the potential functions that were assigned to it; if a node of $T$ has no significant cliques assigned to it then its potential function is taken identically equal to one. After doing so, it performs calculations on $T$ and computes -using a message passing algorithm- the marginal probability distributions of every clique $c \in \mathcal{C}$. The marginal distributions of the significant cliques of the MRF are derived from the marginal distributions of the cliques of set $\mathcal{C}$ by summations. Also, note that by tweaking the algorithm a little it can be used for computing Maximum-A-Posteriori configurations (see for example discussion in [WJ03]).

The single non-standard step of the algorithm is the triangulation of $G$. Different triangulations lead to different clique trees. However, the sizes of the cliques in the clique tree play a crucial role in the running time of the algorithm which is proportional to $\sum_{c \in \mathcal{C}} \prod_{v \in c} |\mathcal{X}_v|$. Note that finding the triangulation that leads to the lightest clique tree under this objective function is an NP-hard optimization problem (see e.g. [BG01] and the references therein) and there are various algorithms for building "good" clique-trees. For an overview of these algorithms see for example [BG01, Bod05]. If $|\mathcal{X}_v| = \chi, \forall v \in V$, then the running time of the algorithm is $O(n \cdot \chi^{width(T)})$, which is at least $O(n \cdot \chi^{treewidth(G)+1})$.

## 3 The Reduction

The reduction that we suggest can be roughly decomposed into two parts, that of translating the graphical game to an appropriately defined Markov Random Field and that of solving a statistical inference problem on the latter. The reduction is the following:

***Frontend***: The Markov Random Field that corresponds to a graphical game $\mathcal{G} = \langle G = (V, E), \{S_p\}, \{u_p\}\rangle$ is defined as follows:
1. The underlying graph of the Markov Random Field is the primal graph $G' = (V', E')$ of $\mathcal{H}(\mathcal{G})$. Since $V' \equiv V$, every node $p \in V'$ corresponds to a unique player in $V$ and we will identify the two.
2. Associated with every node $p \in V'$ is a random variable with state space $S_p$, i.e. a random variable with state space equal to the strategy set of the corresponding player. So $\mathcal{X}_p = S_p$.

3. For every player $p \in V'$, its neighborhood, $\mathcal{N}(p)$, is a clique $c_p$ in $G'$. Note that there might be two players $p_1 \neq p_2$ with $c_{p_1} = c_{p_2}$. The set of significant cliques of the MRF will be the set $\mathcal{C} = \bigcup_{p \in V'} \{c_p\}$. Obviously, $|\mathcal{C}| \leq |V|$.

4. We assign to every player $p$ a function $f_p : \prod_{p' \in c_p} \mathcal{X}_{p'} \to \mathbb{R}_+$ that is defined as follows (call $x(c_p)$ the vector of the random variables corresponding to the players of set $c_p$):

$$f_p(x(c_p)) = \begin{cases} 1, & \text{if } x(c_p)_p \in \mathrm{BR}_{u_p}(x(c_p)_{-p}) \\ \epsilon, & \text{otherwise} \end{cases}$$

where $\epsilon < 1$ is a small constant to be decided later. Intuitively, the function $f_p(x(c_p))$ maps a selection of strategies $x(c_p)$ for the players of the set $\mathcal{N}(p)$ to 1 if the strategy $x(c_p)_p$ of player $p$ is a best response to the strategies $x(c_p)_{-p}$ of his neighbors and to a small constant $\epsilon$ otherwise.

5. We assign to every clique $c \in \mathcal{C}$ a potential function $\psi_c : \prod_{p \in c} \mathcal{X}_p \to \mathbb{R}_+$ defined as follows:

$$\psi_c(x(c)) = \prod_{p \in V' : c_p = c} f_p(x(c))$$

6. Since the Markov Random Field that we defined is parameterized on the choice of $\epsilon$ we will refer to it using the notation $\mathbf{MRF}(\mathcal{G}, \epsilon)$. Also, we will refer to the unnormalized (Z=1) probability distribution defined on $\mathrm{MRF}(\mathcal{G}, \epsilon)$ as $p_\epsilon(x)$; i.e. $p_\epsilon(x) = \prod_{c \in \mathcal{C}} \psi_c(x(c))$.

**Backend**: The computational problems related to pure Nash equilibria are now mapped to statistical inference problems of Markov Random Fields as follows:

a. For any $\epsilon < 1$, finding a MAP configuration, $\hat{x}_{MAP}$, of $\mathrm{MRF}(\mathcal{G}, \epsilon)$ answers the question of whether the graphical game $\mathcal{G}$ has a pure Nash equilibrium. This fact is stated by the following lemmas which are proven in the appendix (call $\mathcal{X} = \prod_v \mathcal{X}_v$).

   **Lemma 3.1** $\forall \epsilon < 1$: ($\mathcal{G}$ has a pure Nash Equilibrium) $\Leftrightarrow$ $(\max_{x \in \mathcal{X}} \{p_\epsilon(x)\} = 1)$

   **Lemma 3.2** $\forall \epsilon < 1$: If $p_\epsilon(\hat{x}) = 1$ for some $\hat{x}$, then $\hat{x} \in \arg\max_{x \in \mathcal{X}} \{p_\epsilon(x)\}$ and $\hat{x}$ is a pure Nash equilibrium of $\mathcal{G}$.

b. Computing the unnormalized marginal distributions[¶] of the significant cliques of $\mathrm{MRF}(\mathcal{G}, 0)$ answers the question of whether the graphical game has a pure Nash equilibrium and, at the same time, the unnormalized marginal distributions constitute a succinct description of all pure Nash equilibria of the graphical game[‖]. These properties of the unnormalized marginal distributions of the significant cliques of the Markov Random Field are stated by the following lemmas which are proven in the appendix.

   **Lemma 3.3** ($\mathcal{G}$ has a pure Nash Equilibrium) $\Leftrightarrow$ ($\forall$ clique $c$ of $\mathrm{MRF}(\mathcal{G}, 0)$, $\exists x^*(c)$ s.t. $p_{0,c}(x^*(c)) \neq 0$), where $p_{0,c}(x(c))$ is the unnormalized marginal probability distribution of clique $c$.

   **Lemma 3.4** If $p_{0,c}(x^*(c)) \neq 0$ for some clique $c$ of $\mathrm{MRF}(\mathcal{G}, 0)$ and some $x^*(c)$ then $\exists$ a pure Nash equilibrium $x^+$ of $\mathcal{G}$ such that $x^+(c) = x^*(c)$.

---

[¶]Those computed assuming Z=1 as discussed in section 2.3.

[‖]Note that there can be exponentially many pure Nash equilibria. A *succinct description* of the set of all pure Nash equilibria (or any other object) $x$ is a string $y$ such that $|y|$ is polynomial in the description of the game and $x = f(y)$ for some function $f$ computable in time polynomial in $|x| + |y|$.

Based on lemmas 3.3 and 3.4 one can build a dynamic programming algorithm that takes as input the marginal probability distributions of all the significant cliques of the MRF and outputs all pure Nash equilibria in output polynomial time.

*Discussion*: We can make the following observations:

1. The reduction described above is completely generic and translates the problem of computing pure Nash equilibria of graphical games to performing statistical inference calculations on Markov Random Fields. The virtue of the reduction is obviously that one can use all the machinery of algorithms developed for statistical inference to attack the problem of computing pure Nash Equilibria; any statistical inference algorithm that computes marginal distributions, maximum-a-posteriori configurations or samples distributions defined on Markov Random Fields works (for now in a heuristic sense) for finding pure Nash equilibria. In sections 4 and 5 we combine this reduction with different statistical inference algorithms and we derive exact deterministic algorithms as well as randomized and heuristic methods for computing pure Nash equilibria.

2. From the proofs of lemmas 3.3 and 3.4 it follows that if a game $\mathcal{G}$ does not have a pure Nash equilibrium then in MRF$(\mathcal{G}, 0)$ function $p_0(x)$ is identically zero. In this case, of course, $p_0(x)$ cannot be a probability distribution of a Markov Random Field. However, this does not affect our use of statistical inference, because statistical inference algorithms are designed to handle this possibility[**]. For example, the algorithms that compute marginal distributions do only summations over the unnormalized probability distribution of the MRF and at a final stage normalize the computed functions if they are not identically zero. Thus, the computations are not affected even if $p_0(x)$ is identically zero.

## 4 Efficient Algorithms For Computing Pure Nash Equilibria

In this section we use our generic reduction to derive polynomial time algorithms for checking the existence of Pure Nash Equilibria and for computing a succinct description of all pure Nash equilibria for large classes of graphical games. The Markov Random Fields to which we will reduce our games will be those with parameter $\epsilon = 0$ and the statistical inference algorithm which we will base our derivations on is the Junction Tree Algorithm described in section 2.4. We note that the algorithms we derive are essentially combinatorial and the following simple observation advocates this. Although the Junction Tree Algorithm performs arithmetic calculations on the potential functions, the only information we really need from the values computed by the Junction Tree algorithm is whether they are zero or positive. Since we start from non-negative entries in our potential functions (the entries are either zeros or ones) and since the Junction Tree algorithm performs no subtractions, one could change the Junction Tree Algorithm arithmetic to account only for whether an entry it computes along the execution is zero or positive. Doing so we need only 1 bit for every stored entry and there are no arithmetic precision issues that we have to address. Thus, our algorithms are essentially combinatorial and our claim that the derived algorithms are polynomial time is valid.

### 4.1 Graphical Games on Trees and Acyclic Hypergraph Games

We will show that graphical games defined on trees and graphical games with acyclic hypergraphs (see [BFMY83] for a definition of hypergraph acyclicity) have efficient pure Nash equilibria computation schemes. Note that the class of games defined on trees is a subclass of the class of games with acyclic hypergraphs as stated by the following easy lemma.

---

[**]Since MRFs are defined in a distributed fashion, i.e. by potential functions on subsets of the nodes, this is an intrinsic problem in statistical inference.

**Lemma 4.1** *Let $\mathcal{G} = \langle G, \{S_p\}, \{u_p\}\rangle$ be a graphical game and let $G$ be a tree. Then $\mathcal{H}(\mathcal{G})$ is acyclic.*

So it is enough to prove our claim for the class of graphical games with acyclic hypergraphs. Before doing so we define the notion of a join tree for a hypergraph and we present a theorem relating *hypergraph acyclicity, Graham's Algorithm and Join Trees*. For a description of Graham's Algorithm we refer to [BFMY83].

**Definition 4.2** A *join tree for a hypergraph* $(\mathcal{P}, \mathcal{H})$ is a tree $T = (V, E)$, where $V \equiv \mathcal{H}$, so that, for all $v_1, v_2 \in V$ and for all $u \in V$ in the unique path between $v_1$ and $v_2$, $v_1 \cap v_2 \subseteq u$.

**Theorem 4.3** *[BFMY83] If $\mathcal{R} = (\mathcal{P}, \mathcal{H})$ is a hypergraph then:*

$$(\mathcal{R} \text{ is acyclic }) \Leftrightarrow (\mathcal{R} \text{ has a join tree }) \Leftrightarrow (\textit{Graham's algorithm succeeds on input } \mathcal{R})$$

We can now prove our claim:

**Theorem 4.4** *Deciding whether a graphical game has a pure Nash equilibrium is in P for all games with acyclic hypergraph. Moreover, computing a succinct description of all pure Nash equilibria can be done in polynomial time.*

**Proof:** On input $\mathcal{G} = \langle G = (V, E), S_p, u_p\rangle$, the algorithm proceeds in the following steps:

1. Apply Graham's Algorithm on $\mathcal{H}(\mathcal{G})$ to check whether it is acyclic; if so, Graham's algorithm returns a join tree $T = (\mathcal{C}, \mathcal{E})$ for $\mathcal{H}(\mathcal{G})$ (for details see [BFMY83]). It is easy to see that $T$ is a clique tree for the primal graph $G'$ of the game.

2. Reduce $\mathcal{G}$ to MRF($\mathcal{G}$,0) as described in section 3. The graph of the MRF is the primal graph of the game and so $T$ is a clique tree for the graph of MRF($\mathcal{G}$,0).

3. Run the Junction Tree Algorithm on $T$ to compute the marginal probability distributions of the significant cliques of MRF($\mathcal{G}$,0).

4. The marginal probability distributions answer the question of whether $\mathcal{G}$ has a pure Nash equilibrium and also constitute a succinct description of all pure Nash equilibria of $\mathcal{G}$ (see section 3)

***Correctness***: The *correctness* of the algorithm follows from the correctness of the reduction and the correctness of Graham's algorithm and the Junction Tree algorithm. Note, also, that since the Junction Tree algorithm maintains *supportiveness* [LS90] it won't be affected by any "divisions by zero".

***Time Complexity***: Graham's algorithm finds the join tree of $\mathcal{H}(\mathcal{G})$ in time polynomial in the size of the hypergraph and, thus, in the number of players. Reducing $\mathcal{G}$ to MRF($\mathcal{G}$,0) also takes polynomial time. It remains to bound the running time of the message-passing phase of the junction tree algorithm. This phase is executed on the clique tree, which is precisely the join tree that Graham's algorithm returned, and it involves the exchange of as many messages as twice the number of edges of the clique tree, so at most $2n - 2$ messages, where $n$ is the number of players (note that the number of nodes of the clique tree is at most equal to the number of players). Now, the time needed to compute a message that is being sent over an edge of the clique tree is polynomial in the size of the tables (potential functions) that are stored at the endpoints of that edge. However, the clique in every node of the clique tree corresponds to the neighborhood of a player and so its table has the same size as the table describing the utility function of that player. Thus, the complexity of every message is polynomial in the input complexity. It is, thus, obvious that the described algorithm runs in time polynomial in the description of the graphical game. ■

## 4.2 Games of Bounded Treewidth and Games of Bounded Hypertree-Width

It is easy to extend the results of section 4.1 to broader classes of graphical games, those of bounded treewidth and bounded hypertree width respectively. The *hypertree width of a graphical game* $\mathcal{G}$ is the hypertree width of $\mathcal{H}(\mathcal{G})$. For a formal definition of the latter we refer the reader to [GLS02], but for quick reference we provide a definition in section A of the appendix. Our results are stated by theorems 4.6 and 4.7. The latter was also proven independently in [GGS03] using different techniques. All proofs of this section are postponed to the appendix. The proof of theorem 4.6 uses lemma 4.5.

**Lemma 4.5** *If the graph $G = (V, E)$ of a graphical game $\mathcal{G}$ has treewidth bounded by $k$ then its primal graph has treewidth bounded by $(k+1) \cdot \max_{p \in V} |\mathcal{N}(p)| - 1$. Moreover, given a tree decomposition of $G$ of width $k$ we can compute in polynomial time a clique tree for the primal graph of $\mathcal{G}$ of width $(k+1) \cdot \max_{p \in V} |\mathcal{N}(p)|$.*

**Theorem 4.6** *Deciding whether a graphical game has a pure Nash equilibrium and computing (a succinct description of) all pure Nash equilibria is in P for all classes of games with bounded treewidth.*

**Theorem 4.7** *Deciding whether a graphical game has a pure Nash equilibrium and computing (a succinct description of) all pure Nash equilibria is in P for all classes of games with bounded-hypertreewidth.*

## 4.3 Games of $O(\log n)$-Treewidth

The classes of graphical games for which we have derived efficient pure Nash equilibria computation schemes are quite broad. However, often in games of practical interest the treewidth and hypertree width of the underlying graph are not bounded. In this section we go one step further in our study of graphical games and study classes of games with $O(\log n)$ treewidth. For these classes, we derive polynomial time pure Nash equilibria computation algorithms under the assumption of bounded neighborhood size and bounded cardinality strategy sets. Given that NP-completeness of computing pure Nash equilibria holds even when restricted to the class of graphical games with bipartite graphs of degree at most 3 and strategy sets of cardinality at most 3, our result is quite tight. Moreover, it is the first positive result for computing pure Nash equilibria that is not based on some assumption about the cycle structure of the graph. For a different setting for studying big games where succinct description is required see [DP05]. Also, see theorem 4.9 for a relaxation of the bounded neighborhood requirement.

**Theorem 4.8** *Deciding whether a graphical game has a pure Nash equilibrium is in P for all classes of games with $O(\log n)$ treewidth, bounded cardinality strategy sets and bounded neighborhood size. Moreover, computing a succinct description of all pure Nash equilibria can be done in polynomial time.*
*Proof:* The algorithm is similar in spirit to the ones presented so far, but differs in the construction of the clique tree on which the Junction Tree algorithm performs. This task is somewhat involved and is described here. Suppose $k = k(n)$; by slightly modifying the algorithm presented by Becker and Geiger [BG01], we get an algorithm that runs in time $poly(n) \cdot 2^{4.67 \cdot k}$, on input a graph $G$ of $n$ nodes, and either outputs a tree decomposition of $G$ of width at most $3.67k$ or outputs that the treewidth of $G$ is larger than $k$[††]. For $k = c \log n$, where $c$ is a fixed constant, we get an algorithm that runs in time polynomial in $n$ and either returns a tree decomposition of the input graph $G$ of width at most $3.67c \log n$ or outputs that the treewidth of $G$ is larger than $c \log n$. Now, suppose $\mathcal{G} = \langle G, \{S_p\}, \{u_p\} \rangle$ is a game drawn from a family of games with treewidth at most $c \log n$. Applied to $G$, the algorithm returns a tree decomposition of $G$ of width at most $3.67 \cdot c \cdot \log n$. Given this tree decomposition, from lemma 4.5, we can construct in polynomial time a clique tree for the primal graph of $\mathcal{G}$, which is the graph of the MRF, of width $w = (3.67 \cdot c \cdot log n + 1) \cdot \max_{p \in V} |\mathcal{N}(p)|$. Now, if we assume bounded cardinality strategy sets, it follows that the sizes of the tables (potential functions) that will be stored in the clique tree before the execution of the junction tree algorithm and, thus, all the messages exchanged during the execution have size $O\left(n^{O(\max |\mathcal{N}(p)|)}\right)$. If, moreover, we assume bounded neighborhood size they are polynomial in the number of players. So all the computation takes polynomial time. ∎

---
[††]Alternatively we could use Reed's approximation algorithm [Ree92] for treewidth or other approximation algorithms.

We can get rid of the bounded neighborhood requirement by pushing the $O(\log n)$-treewidth requirement to the primal graph of the game as stated by the following theorem which is proven in the appendix. This way we can in some cases accommodate neighborhoods of size up to $O(\log n)$ which might be helpful in some applications.

**Theorem 4.9** *Deciding whether a graphical game has a pure Nash equilibrium is in P for all classes of games with primal graphs of treewidth $O(\log n)$ and bounded cardinality strategy sets. Moreover, computing a succinct description of all pure Nash equilibria can be done in polynomial time.*

## 5 Further

### 5.1 A General Algorithmic Scheme

In section 4 we combined our reduction with the junction tree algorithm and derived polynomial time algorithms for computing pure Nash equilibria for large classes of graphical games. The high level schema of our algorithms is the following:

- Reduce input game $\mathcal{G}$ to MRF($\mathcal{G}, 0$).
- Find a good clique tree of the MRF graph.
- Run the junction tree algorithm on the clique tree.

In essence, our algorithms try to reconcile two things. On the one hand, the running time of the junction tree algorithm crucially depends on the width of the clique tree compared to the size of the largest neighborhood of the game. On the other hand, computing the optimal clique tree is an NP-hard problem. To circumvent this difficulty we can make the following observation. We do not really need to find the optimal clique tree. To preserve efficiency, a constant factor approximation is enough: if the junction tree algorithm on the optimal clique tree runs in polynomial time, then it runs in polynomial time on a constant factor approximation of the optimal tree. Moreover, the running time of the junction tree algorithm will be polynomial in $n \cdot s^w$, where $s$ is the number of strategies and $w$ is the width of the tree decomposition we achieve. So, an algorithm for computing a constant factor approximation to the treewidth is sufficiently fast if its running time is polynomial in $n \cdot s^w$, where $w$ is the constant factor approximation to the treewidth that it achieves. In fact, there are various approximation algorithms for treewidth that have this property. The one given in [BG01] runs in time proportional to $poly(n) \cdot 2^{4.67 \cdot k}$, where $k$ is the treewidth, and returns a clique tree of width at most $3.67 \cdot k$; also Reed's algorithm [Ree92], Robertson and Seymour's algorithm [RS95] and other algorithms [Bod05] have this property. Therefore, we have optimal algorithms under the above scheme. Notably, *all algorithms presented in section 4 can be derived as special cases of this scheme*. Thus, our general algorithm encompasses all the positive results for computing pure Nash equilibria known to date plus the new positive results discovered in this paper.

### 5.2 Heuristics

The use of the junction tree algorithm provided the deterministic polynomial time algorithms presented in the previous sections. However, our reduction permits the use of a large variety of statistical inference algo-rithms to attack our problem. In this section we comment on two families of statistical inference algorithms that we believe are promising in solving graphical games: Markov Chain Monte Carlo methods and Survey Propagation.

### 5.2.1 Sampling and Simulated Annealing

Our definition of MRF$(\mathcal{G}, \epsilon)$ ensures that for some small $\epsilon(\mathcal{G}) < 1$ most of the probability mass is concentrated on the set of pure Nash equilibria of $\mathcal{G}$, if equilibria exist. More generally, the probability of a configuration decays exponentially in the number of unsatisfied players. This observation suggests quite naturally the use of sampling techniques to find pure Nash equilibria of games or approximations to pure Nash equilibria, i.e. configurations with as many satisfied players as possible. Markov Chain Monte Carlo methods and Simulated Annealing (see e.g. [GRS96]) can be easily applied for computing pure Nash equilibria under our reduction.

### 5.2.2 Survey Propagation Algorithms for solving Hard Instances

An algorithm motivated by statistical inference that appears to be very effective in solving random $k$-SAT instances and other constraint satisfaction problems is survey propagation (e.g. [BMZ02]). Survey propagation performs better than belief propagation for $k$-SAT instances near the SAT/UNSAT threshold (see e.g. [AP03] for the later). In [MMW05] survey propagation is extended to a family of survey propagation algorithms (parameterized by a real number $\rho \in [0, 1]$) that has at one extreme ($\rho = 0$) belief propagation and on the other extreme ($\rho = 1$) survey propagation. By reducing graphical games to Markov Random Fields we can use all algorithms of this family to compute pure Nash equilibria. We believe that survey propagation algorithms for pure Nash equilibria will be very useful for solving hard instances of graphical games.

## Acknowledgements

I would like to thank Christos Papadimitriou and Elchanan Mossel for insightful remarks during the preparation of the manuscript.

# APPENDIX

## A  Hypertree Decompositions and Hypertree width

For quick reference, we define here the notion of a *hypertree decomposition* and *hypertree-width* of a hypergraph. For more details on the properties of hypertree-width and its relation to other notions of hypergraph acyclicity we refer the reader to [GLS02].

**Definition A.1 ([GLS02])** Let $\mathcal{R} = (\mathcal{N}, \mathcal{H})$ be a hypergraph. A *hypertree decomposition* of $\mathcal{R}$ is a triplet $\langle T, \chi, \lambda \rangle$, where $T = (V, E)$ is a rooted tree and $\chi, \lambda$ are labelling functions associating each vertex $v \in V$ with two sets $\chi(v) \subseteq \mathcal{N}$ and $\lambda(v) \subseteq \mathcal{H}$, so that:

1. $\forall h \in \mathcal{H}, \exists v \in V : h \subseteq \chi(v)$
2. $\forall n \in \mathcal{N}$, the set $\{v \in V | n \in \chi(v)\}$ induces a connected subgraph of $T$
3. $\forall v \in V, \chi(v) \subseteq \bigcup_{h \in \lambda(v)} h$
4. $\forall v \in V, \chi(T_v) \cap \bigcup_{h \in \lambda(v)} h \subseteq \chi(v)$
   where $T_v$ is the subtree of $T$ rooted at $v$ and $\chi(T_v) = \bigcup_{v' \in vert(T_v)} \chi(v')$

The *width* of $(T, \chi, \lambda)$ is $max_{v \in V}\{|\lambda(v)|\}$.

The *hypertree width* of a hypergraph $\mathcal{R}$, $hw(\mathcal{R})$, is the minimum width over all its hypertree decompositions.

## B  Missing Proofs

*Proof of lemma 3.1:*
($\Rightarrow$) We have:

$$\begin{aligned}
&x \text{ is a pure Nash equilibrium of } \mathcal{G} \\
&\Rightarrow \forall p \in V' : x(c_p)_p \in \text{BR}_{u_p}(x(c_p)_{-p}) \\
&\Rightarrow \forall p \in V' : f_p(x(c_p)) = 1 \\
&\Rightarrow \forall c \in \mathcal{C} : \psi_c(x(c)) = 1 \\
&\Rightarrow p_\epsilon(x) = 1 \\
&\Rightarrow x \in \arg\max_{x \in \mathcal{X}} \{p_\epsilon(x)\} \text{ ( since by definition } p_\epsilon(x) \leq 1, \forall x) \\
&\Rightarrow \max_{x \in \mathcal{X}} p_\epsilon(x) = 1
\end{aligned}$$

($\Leftarrow$) We have:

$$\mathcal{G} \text{ has no pure Nash equilibria} \Rightarrow$$
$$\nexists x \in \mathcal{X} \text{ s.t. } \forall p \in V' : x(c_p)_p \in \text{BR}_{u_p}(x(c_p)_{-p}) \Rightarrow$$
$$\nexists x \in \mathcal{X} \text{ s.t. } \forall p \in V' : f_p(x(c_p)) = 1 \stackrel{\epsilon \leq 1}{\Rightarrow}$$
$$\nexists x \in \mathcal{X} \text{ s.t. } \forall c \in \mathcal{C} : \psi_c(x(c)) = 1 \Rightarrow$$
$$\nexists x \in \mathcal{X} \text{ s.t. } p_\epsilon(x) = 1 \Rightarrow$$
$$\max_{x \in \mathcal{X}} p_\epsilon(x) < 1 \text{ ( since by definition } p_\epsilon(x) \leq 1, \forall x)$$

∎

*Proof of lemma 3.2:* If $p_\epsilon(\hat{x}) = 1$ then $\hat{x} \in \arg\max_{x \in \mathcal{X}} p_\epsilon(x)$ since $p_\epsilon(x) \leq 1$ by definition ($\epsilon < 1$). Also:

$$p_\epsilon(\hat{x}) = 1 \stackrel{\epsilon \leq 1}{\Leftrightarrow}$$
$$\forall c \in \mathcal{C} : \psi_c(\hat{x}(c)) = 1 \stackrel{\epsilon \leq 1}{\Leftrightarrow}$$
$$\forall p \in V' : f_p(\hat{x}(c_p)) = 1 \Leftrightarrow$$
$$\forall p \in V' : \hat{x}(c_p)_p \in \text{BR}_{u_p}(\hat{x}(c_p)_{-p}) \Leftrightarrow$$
$$\hat{x} \text{ is a pure Nash equilibrium}$$

∎

*Proof of lemma 3.3:*
Note that the marginalization of $p_0(x)$ with respect to a clique $c$ of MRF$(\mathcal{G}, 0)$ is simply a summation $p_{0,c}(x(c)) = \sum_{x_v, v \notin c} p_0(x)$. Now it is easy to prove the claim as follows:
($\Rightarrow$) If $\mathcal{G}$ has a pure Nash equilibrium $x^*$ then $p_0(x^*) = 1$ (see proof of lemma 3.2). It follows that $p_{0,c}(x^*(c)) > 0$ for every clique $c$ of the MRF.
($\Leftarrow$) Proof by contradiction:

$$\mathcal{G} \text{ does not have a pure Nash equilibrium}$$
$$\Rightarrow \forall x, \exists p \text{ s.t. } f_p(x(c_p)) = \epsilon$$
$$\stackrel{\epsilon=0}{\Rightarrow} \forall x, p_0(x) = 0$$
$$\Rightarrow \forall \text{ clique } c, \forall x(c) : p_{0,c}(x(c)) = 0$$

∎

*Proof of lemma 3.4:*
Proof by contradiction:

$$\text{There is no pure Nash equilibrium } x^+ \text{ of } \mathcal{G} \text{ such that } x^+(c) = x^*(c)$$
$$\Rightarrow \forall x \text{ with } x(c) = x^*(c) : p_0(x) = 0$$
$$\Rightarrow p_{0,c}(x^*(c)) = \sum_{x:x(c)=x^*(c)} p_0(x) = 0$$

∎

*Proof of lemma 4.5:* (sketch) Let $T = (\mathcal{C}, \mathcal{E})$ be a tree decomposition of the game graph $G = (V, E)$, where $\mathcal{C} \subseteq 2^V$, and suppose that $k = \max_{c \in \mathcal{C}} |c| - 1$ is the width of the decomposition. We show how to construct a tree decomposition $T'$ of the primal graph of $\mathcal{G}$ of width at most $(k+1) \cdot \max_{p \in V} |\mathcal{N}(p)| - 1$. $T'$ is isomorphic to $T$ and let $\sigma$ denote the one-to-one correspondence between vertices of $T$ and $T'$. Then for all $c \in \mathcal{C}$ we set $\sigma(c) = \bigcup_{p \in c} \mathcal{N}(p)$, i.e. every vertex of $T'$ contains the union of the neighborhoods of all players of the corresponding vertex of $T$. It is not difficult to see that $T'$ is a tree decomposition of the primal graph $G' = (V, E')$ of the game. Indeed, for every edge $(u, v) \in E'$ there is a vertex of $T'$ that contains both $u, v$: if $(u, v)$ is an edge in $G'$ then players $u, v$ must belong in the neighborhood of some player $p$ (maybe $p \equiv u$ or $p \equiv v$); but there is at least one vertex of $T$ that contains $p$ and, thus, the corresponding vertex of $T'$ must contain all its neighborhood and, so, $u, v$ as well. Moreover, for every player $p \in V$ the vertices of $T'$ that contain $p$ form a connected subtree of $T'$: since $T$ is a tree decomposition of $G$, it is easy to see that the vertices of $T$ that contain player $p$ or a neighboring player of $p$ in $G$ form a connected component in $T$; but, in $T'$, $p$ appears in exactly those vertices whose corresponding vertex in $T$ contains either $p$ or a neighbor of $p$ in $G$, and, so, the nodes in which $p$ appears must form a connected component in $T'$. Finally, since $k + 1 = \max_{c \in \mathcal{C}} |c|$, every vertex of $T'$ contains at most $(k+1) \cdot \max_{p \in V} |\mathcal{N}(p)|$ vertices. Given $T$, the construction of $T'$ can be done in polynomial time. ∎

*Proof of theorem 4.6:* For a fixed constant $k$ the algorithm performs the following steps on input $\mathcal{G} = \langle G = (V, E), \{S_p\}, \{u_p\}\rangle$:

1. Check whether $G$ has treewidth bounded by $k$ and, if so, find a tree decomposition of $G$ of width at most $k$. For details on how to perform this step in polynomial time see for example [ACP87, Klo94].

2. From the tree decomposition of $G$ get a clique tree $T'$ of the primal graph of $\mathcal{G}$ that has width at most $(k+1) \cdot \max_{p \in V} |\mathcal{N}(p)|$ using lemma 4.5.

3. Reduce $\mathcal{G}$ to MRF($\mathcal{G}$,0) as described in section 3. The graph of the MRF is the primal graph of the game and so $T'$ is a clique tree for the graph of MRF($\mathcal{G}$,0).

4. Run the Junction Tree Algorithm on $T'$ to compute the marginal probability distributions of the significant cliques of MRF($\mathcal{G}$,0). The marginal probability distributions answer the question of whether $\mathcal{G}$ has a pure Nash equilibrium and also constitute a succinct description of all pure Nash equilibria of $\mathcal{G}$.

The *correctness* of the algorithm follows easily from the correctness of every intermediate step. The *running time* of the algorithm is polynomial. This is easily proven using the same rationale as in the proof of theorem 4.4. However, in this case the cliques contained in the nodes of the clique tree have size that is at most $(k+1) \cdot \max_{p \in V} |\mathcal{N}(p)|$, i.e. at most $k+1$ times the size of the biggest neighborhood of the game. Since the biggest table of the input has dimension $\max_{p \in V} |\mathcal{N}(p)|$ and the biggest table in the clique tree has dimension $k+1$ times $\max_{p \in V} |\mathcal{N}(p)|$, where $k$ is fixed, it follows that the tables of the clique tree have size polynomial in the input complexity. Thus, the algorithm runs in polynomial time. ∎

*Proof of theorem 4.7:* The proof is based on the following graph-theoretic lemma.

**Lemma B.1** *If a graphical game $\mathcal{G} = \langle G = (V, E), \{S_p\}, \{u_p\}\rangle$ has hypertree width bounded by $k$ then its primal graph has treewidth bounded by $k \cdot \max_{p \in V} |\mathcal{N}(p)| - 1$. Moreover, given a hypertree decomposition $(T, \chi, \lambda)$ of $\mathcal{H}(\mathcal{G})$ we can compute in polynomial time a clique tree for the primal graph of $\mathcal{G}$ of width $k \cdot \max_{p \in V} |\mathcal{N}(p)|$, where $k$ is the width of $(T, \chi, \lambda)$.*

**Proof:** Given a hypertree decomposition $(T, \chi, \lambda)$ of $\mathcal{H}(\mathcal{G})$, take $T'$ to be a tree isomorphic to $T$ after removing directions from the edges and let $\sigma$ denote the one-to-one correspondence between vertices of $T$ and $T'$. Then for all $v \in vert(T)$ we set $\sigma(v) = \chi(v)$. It is easy to see that $T'$ is a clique tree for the primal graph of $\mathcal{G}$, using properties 1 and 2 of the hypertree decomposition (see definition A.1). Moreover, from property 3 it follows that for every node $v$ of $T$: $\chi(v) \subseteq \bigcup_{h \in \lambda(v)} h$. Thus, for every node $v$ of $T$: $|\chi(v)| \leq k \cdot \max_{p \in V} |\mathcal{N}(p)|$, where $k$ is the width of the hypertree decomposition $(T, \chi, \lambda)$. Thus, the clique tree $T'$ has width at most $k \cdot \max_{p \in V} |\mathcal{N}(p)|$. The construction can obviously be done in polynomial time. ∎

For a fixed constant $k$ the algorithm performs the following steps on input $\mathcal{G} = \langle G = (V, E), \{S_p\}, \{u_p\}\rangle$:
1. Check whether $\mathcal{H}(\mathcal{G})$ has hypertree width bounded by $k$; if so find a hypertree decomposition $(T, \chi, \lambda)$ of $\mathcal{H}(\mathcal{G})$ of width at most $k$. For details on how to perform this step in polynomial time see [GLS02].

2. From the hypertree decomposition $(T, \chi, \lambda)$ get a clique tree $T'$ of the primal graph of $\mathcal{G}$ that has width at most $k \cdot \max_{p \in V} |\mathcal{N}(p)|$ using lemma B.1.

3. run algorithm of theorem 4.6 from step 3;

The *correctness* and the *time complexity* of the algorithm are analyzed in the same way as in the proof of theorem 4.6. ∎

*Proof of theorem 4.9:* The algorithm is similar in spirit to the one presented in the proof of theorem 4.8. Again, we use the modified -as described in the proof of theorem 4.8- algorithm of Becker and Geiger for $k = c \log n$, where $c$ is a fixed constant, but we apply it directly on the graph $G$ of the MRF. The algorithm runs in polynomial time and if $G$ has treewidth bounded by $c \log n$, it returns a tree decomposition of $G$ of width at most $3.67 c \log n$. So the biggest table in the clique tree will be of dimension $3.67 c \log n$. Thus, assuming bounded cardinality strategy sets, the sizes of the tables (potential functions) that we store at the nodes of the clique tree are polynomial in the number of players so all the computation takes polynomial time. ∎